\documentclass[%
 reprint,
superscriptaddress,
 amsmath,amssymb,
 aps,
 twocolumn,longbibliography
]{revtex4-1}

\usepackage[colorlinks=true,urlcolor=blue,citecolor=black,linkcolor=black]{hyperref}
\usepackage{times}
\usepackage{color}
\usepackage{xspace}
\usepackage{graphicx}
\usepackage{dcolumn}

\newcommand{\fig}{Fig.~}
\newcommand{\eq}{Eq.~}

\newcommand{\vb}[1]{ {\bf #1}}

\newcommand{\rb}{\ensuremath{\vb{r}}\xspace}
\newcommand{\cim}{\ensuremath{\overline{\textrm{CI}}}\xspace}
\newcommand{\snrm}{\ensuremath{\overline{\textrm{SNR}}}\xspace}
\newcommand{\resp}[1]{#1}
\begin{document}


\title{Chemotaxis in uncertain environments: hedging bets with multiple receptor types}

\author{Austin Hopkins}
\affiliation{%
 Department of Physics \& Astronomy, Johns Hopkins University
}
\author{Brian A. Camley}%
\affiliation{%
 Department of Physics \& Astronomy, Department of Biophysics, Johns Hopkins University
}
%


\begin{abstract}
Eukaryotic cells are able to sense chemical gradients in a wide range of environments. We show that, if a cell is exposed to a highly variable environment, it may gain chemotactic accuracy by expressing multiple receptor types with varying affinities for the same signal, as found commonly in chemotaxing cells like Dictyostelium. As environment uncertainty is increased, there is a transition between cells preferring a single receptor type and a mixture of types -- hedging their bets against the possibility of an unfavorable environment. We predict the optimal receptor affinities given a particular environment. In chemotaxing, cells may also integrate their measurement over time. Surprisingly,  time-integration with multiple receptor types is qualitatively different from gradient sensing by a single type -- cells may extract orders of magnitude more chemotactic information than expected by naive time integration. Our results show when cells should express multiple receptor types to chemotax, and how cells can efficiently interpret the data from these receptors.
\end{abstract}

\maketitle

As a white blood cell finds a wound, or an amoeba finds nutrition, they chemotax, sensing and following chemical gradients. Eukaryotic cells sense gradients of chemical ligands by measuring ligand binding to receptors on the cell's surface. In eukaryotic chemotaxis in shallow gradients, accuracy is limited by unavoidable stochasticity arising from randomness in ligand-receptor binding and diffusion \cite{levine2013physics,fuller2010external,ueda2007stochastic,segota2013high}. Eukaryotic chemotaxis has been extensively modeled \cite{levine2013physics,hu2010physical,endres2008accuracy,shi2013interaction,hecht2011activated}, including extensions to collective chemotaxis \cite{camley2018collective,camley2017cell,mugler2016limits,hopkins2019leader} and stochastic simulation \cite{sharma2016gradient,lakhani2017testing}. Modeling and experiment show eukaryotic chemotaxis is most accurate at ligand concentrations near the receptor dissociation constant $K_D$ \cite{fuller2010external,hu2010physical,ueda2007stochastic}. 
Cells crawling through tissue and searching for targets at variable concentrations are exposed to a huge variation in environmental signals. \resp{Cells often express multiple receptors for the same signal, with $K_D$ values ranging over orders of magnitude \cite{de1985binding,johnson1992cyclic}. For instance, during Dictyostelium's life cycle, Dicty expresses multiple different combinations of cAMP receptors CAR1-CAR4 \cite{hereld1993camp}, with ranges of $K_D$ from 25 nM to $>5000$ nM. Larger-$K_D$ receptors are expressed later in development, when the cAMP background level rises; the change in receptor expression has been suggested to allow Dicty to deal with the new environments \cite{kim1998switching,dormann2001camp}. In addition, Segota et al. recently showed that to explain the high accuracy of Dictyostelium chemotaxis to folic acid over a broad range of folic acid concentrations, multiple receptor types (with $K_D$ values ranging from 2 nM to 450 nM \cite{de1985binding}) and multiple measurements over time must be accounted for \cite{segota2013high}.}

We argue that if a cell is sufficiently uncertain about its chemical environment, it should express multiple receptor types. We provide results for the optimal receptor $K_D$s depending on environmental uncertainty. In addition, we show that integrating information from multiple measurements of the receptor binding state is more complicated in the many-receptor-type case, and show that a standard approach significantly under-estimates gradient sensing accuracy.

We generalize the model of \cite{hu2010physical,hu2011geometry}, considering a cell with receptors of $R$ types with dissociation constants $K_D^i$, $i = 1\cdots R$, with receptors spread evenly over a circular cell. We find the fundamental limit set by the Cram\'er-Rao bound with which this cell can measure a shallow gradient $\vb{g}$ using a snapshot of current receptor occupation. If the concentration near the cell is locally $c(\rb) = c_0 e^{\rb\cdot\vb{g} / L}$, i.e. g is the percentage change across the cell diameter $L$, this uncertainty is (Appendix \ref{app:snapshot}):
\begin{equation}
    \sigma_\vb{g}^2 = \sigma_{g_x}^2 + \sigma_{g_y}^2 = \frac{16}{N \sum_{i=1}^R f_i \frac{c_0 K_D^i}{ \left(c_0 + K_D^i\right)^{2}}} \; \; \; \textrm{(snapshot)} \label{eq:variance_snapshot}
\end{equation}
where $N$ is the total receptor number, and $f_i$ is the fraction of receptors that are type $i$.

\begin{figure}[h!]
    \centering
   \includegraphics[width=\linewidth]{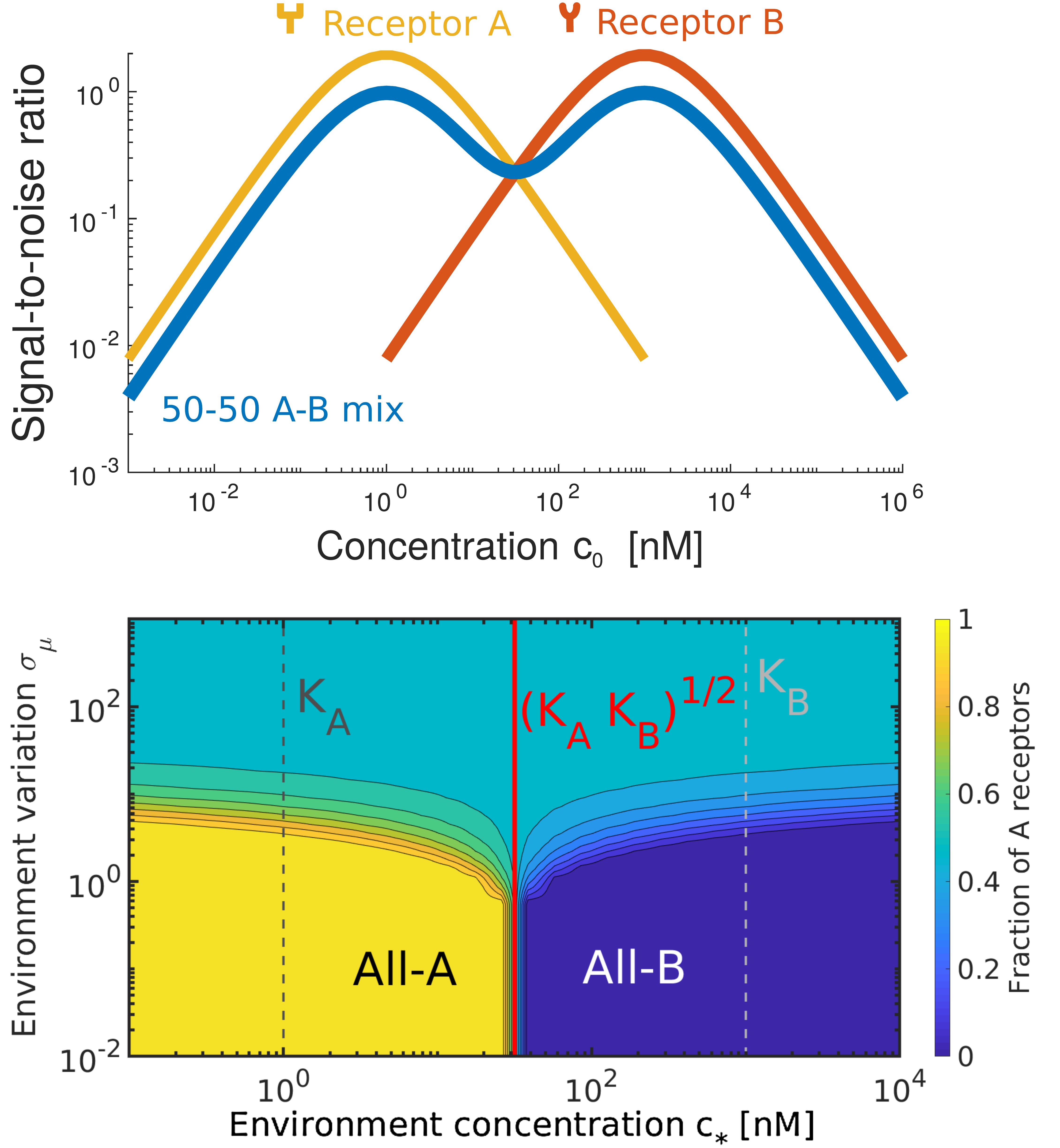}
    \caption{\textbf{Cells benefit from mixed receptor expression when their environment is uncertain---} TOP: The signal-to-noise ratio $g^2/\sigma_g^2$ from \eq \ref{eq:variance_snapshot} for 100\% $A$ receptors (yellow), 100\% $B$ receptors (red) or both in a 50-50 combination (blue).  BOTTOM: Fraction of $A$ receptors that maximizes \cim for snapshot measurements (\eq \ref{eq:variance_snapshot}) as a function of $p(c_0)$ where $p(\ln c_0) \sim e^{-(\ln c_0-\ln c_*)^2/2\sigma_\mu^2}$, characterized by the environment concentration $c_*$ and the standard deviation of the log of the concentration, $\sigma_\mu$. The transition between All-$A$ and All-$B$ occurs at $\sqrt{K_A K_B}$ at low uncertainty. Dashed lines indicate receptor $K_D$. $K_A = 1$ nM, $K_B = 1000$ nM, $N = 5 \times 10^4$ and $g = 0.05$ in both plots. }
    \label{fig:hedge1}
\end{figure}

When can a cell improve its accuracy by expressing multiple receptor types? If we try to maximize  signal-to-noise ratio $\textrm{SNR} = g^2/\sigma_g^2$, we see from \eq \ref{eq:variance_snapshot} that SNR is a linear function of the $f_i$, and will be maximized by choosing $f = 1$ for the type with the largest value of $\frac{c_0 K_D^i}{(c_0+K_D^i)^2}$, and $f = 0$ for the others. For the simplest case of two types $A$ and $B$ with dissociation constants $K_A, K_B$ ($K_B > K_A$) this means that the best accuracy occurs with all $A$ receptors when $c_0 < \sqrt{K_A K_B}$, and with all $B$ receptors when $c_0 > \sqrt{K_A K_B}$ (\fig \ref{fig:hedge1}, top). 

If the background concentration is completely known, it is never beneficial for a cell to express multiple receptor types simultaneously. However, if a cell is uncertain about the concentration it is likely to encounter, it may hedge its bets by expressing multiple receptor types, allowing it to chemotax effectively in more environments. Does a cell in concentration $c_0$ with probability $p(c_0)$ benefit from multiple receptor types? What metric is appropriate? We could compute average signal-to-noise ratio, $\snrm = \int d c_0 p(c_0) g^2/\sigma_g^2$, but increasing SNR at one concentration is little consolation to the cell that finds itself completely lost at another concentration -- the utility of SNR saturates. We therefore optimize the mean chemotactic index \cim, where $\textrm{CI} = f(\textrm{SNR})$, with $f(x)$ saturating at 1 as $x\to\infty$. Here, we use $f(x) = \sqrt{2 x / \pi} / L_{1/2}(-x/2)$, where $L_{1/2}(x)$ is a generalized Laguerre polynomial, as in \cite{camley2016emergent};  alternate definitions of CI lead to similar results. 

In \fig \ref{fig:hedge1}, we consider two receptor types with dissociation constants $K_A$ and $K_B$, and numerically determine the fraction of $A$ receptors that maximizes \cim. We choose $p(c_0)$ to be log-normal, $p(\ln c_0) \sim e^{-(\ln c_0 - \ln c_*)^2/2 \sigma_\mu^2}$ -- a generic option for large variability. When environmental uncertainty $\sigma_\mu$ is small, we see the behavior predicted above -- at small $c_*$, the cell should express all $A$ receptors, while for $c_* > \sqrt{K_A K_B}$ the cell switches to all-$B$. However, at larger $\sigma_\mu$, cells optimize \cim by expressing equal amounts of $A$ and $B$ receptors (\fig \ref{fig:hedge1}, bottom). 

Why is a 50-50 mix optimal even when $c_* \approx K_A$? This may seem like a natural response to uncertainty, but it is not obvious why, if the typical concentration, $c_*$ is close to $K_A$, the cell would not prefer $A$-type receptors. We argue that in the limit of large $\sigma_\mu$, where $p(c_0)$ becomes very broad, but $\textrm{CI}(c_0)$ is locally peaked, the optimal fraction should not depend on $c_*$ or $\sigma_\mu$.
The chemotactic index $\textrm{CI}(c_0)$ for snapshot sensing is generally peaked when $c_0$ is around $K_A$ or $K_B$, because the SNR decays when $c_0 \ll K_A$ or $c_0 \gg K_B$ (Fig. \ref{fig:hedge1}, top). As $\sigma_\mu$ increases and $p(c_0)$ becomes more weakly dependent on $c_0$, we can approximate the integral defining $\cim$ as 
\begin{align}
\nonumber\cim &= \int d c_0 p(c_0) \textrm{CI}(c_0) \\
    &\approx p(\sqrt{K_A K_B}\,) \int dc_0 \textrm{CI}(c_0)
    \label{eq:5050approx}
\end{align}
(We've chosen $\sqrt{K_A K_B}$ as a typical value in the range $K_A\cdots K_B$.) In \eq \ref{eq:5050approx}, the parameters  $c_*$ and $\sigma_\mu$ of the environment distribution $p(c_0)$ only appear in $p(\sqrt{K_A K_B})$, and the receptor fractions only appear in the term $\int dc_0 \textrm{CI}(c_0)$. In this limit,  $p(\sqrt{K_A K_B})$ becomes an irrelevant prefactor -- the same fraction will optimize \cim independent of $c_*$ and $\sigma_\mu$, and so we see a 50-50 mix for a broad range of parameters. We will see later that the 50-50 mix is no longer optimal when cells time average and $\textrm{CI}(c_0)$ is no longer locally peaked.

The 50-50 mix between $A$ and $B$ receptors in \fig \ref{fig:hedge1} emerges when $p(c_0)$ is so broad it is slowly-varying on the scale of $\textrm{CI}(c_0)$. We caution that at these large levels of environmental variation, the difference between the optimal receptor configuration and simply choosing all-$A$ or all-$B$ receptors is small (Appendix \ref{app:extended}) -- at sufficiently high uncertainties, no configuration is particularly successful.

\begin{figure}[tpb]
    \centering
    \includegraphics[width=\linewidth]{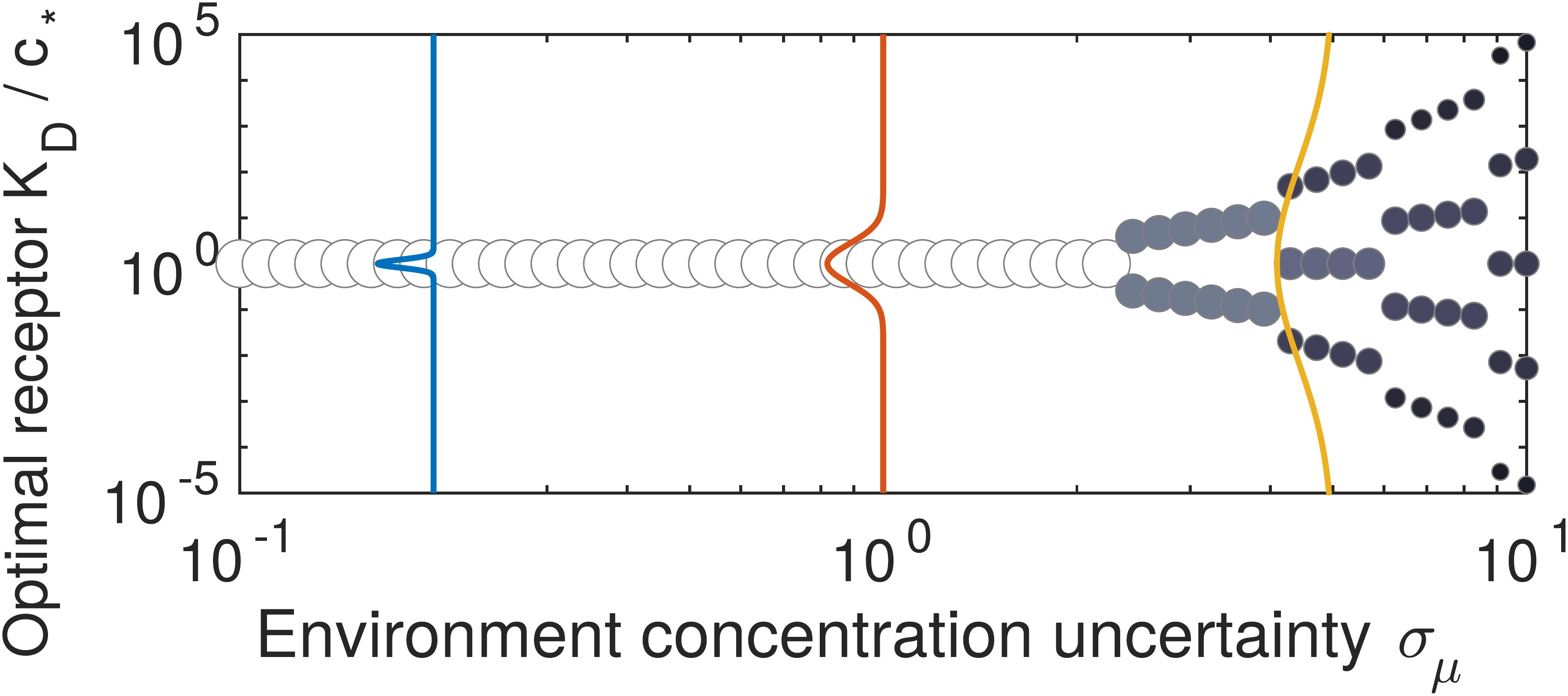}
    \caption{\textbf{Optimal receptor configuration.} Best $K_D^i$ values for receptors as a function of environmental uncertainty $\sigma_\mu$. Marker areas are scaled to the fraction of that receptor type.
    Solid lines illustrate $p(\ln c_0)$. $c_* = 1$ nM, $N = 5\times10^4$ and $g = 0.05$, $\Delta_{CI} = 0.01$ and a maximum of $R = 7$ types are considered.}
    \label{fig:branch}
\end{figure}

\fig \ref{fig:hedge1} shows when a cell should choose to express a combination of receptor types. Can we also find which receptors a cell would evolve to maximize  gradient-sensing ability in a given $p(c_0)$? We optimize \cim by varying $K_D^i$ and receptor fraction $f_i$ for different numbers of receptor types $R$ and different widths $\sigma_\mu$, holding total receptor number $N$ constant. We choose the configuration that maximizes \cim -- with an important caveat. By adding more types with arbitrary $K_D$, we can always at least match the performance of a single type. If many configurations generate roughly the same near-optimal $\cim$ (all within $\Delta_{CI} = 0.01$), we choose from these the configuration with the fewest receptor types $R$. To reduce the number of variables we vary, we use the symmetry of $p(c_0)$, assuming $\ln K_D^i$ and $f_i$ are mirror-symmetric around $\ln c^*$. The resulting optimal $K_D^i$ are shown in \fig \ref{fig:branch}. We see that as the environment uncertainty $\sigma_\mu$ increases, there is a transition between preferring a single receptor type and multiple receptor types, with the $K_D$ values for the multiple types being spread over the likely range of concentrations observed. 

\begin{figure}[ht]
    \centering
    \includegraphics[width=\linewidth]{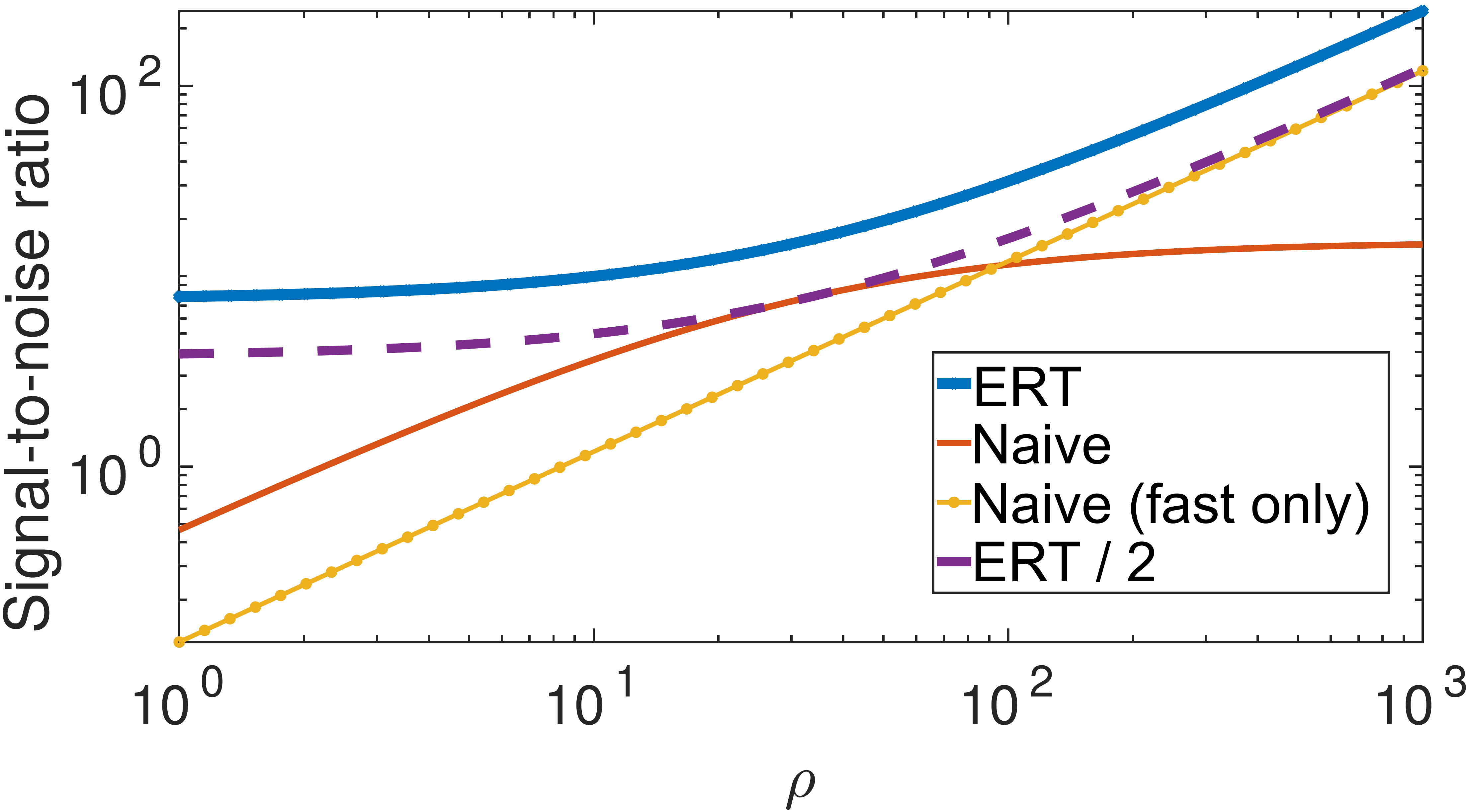}
    \caption{\textbf{Estimation using the entire receptor trajectory is much more accurate than naive averaging.---} SNR $g^2/\sigma_g^2$ as a function of $\rho = k_-^B / k_-^A$; $\rho > 1$ states $B$ receptors have faster off rates. SNR is much larger using the entire receptor trajectory (ERT) method; even in its best case, the naive average only reaches half of the ERT SNR. We use two receptor types, $K_A = 1$ nM, $K_B = 1000$ nM, and $N = 5 \times 10^4$, $g = 0.05$, with fixed $c_0 = \sqrt{K_A K_B}$ and $f_A = 0.5$, with $k_-^A T = 2.$}
    \label{fig:SNR}
\end{figure}

\fig \ref{fig:hedge1} and \fig \ref{fig:branch} are based on \eq \ref{eq:variance_snapshot}, which gives the fundamental uncertainty for a cell sensing a gradient {\it from a single snapshot of its receptors.} If cells integrate measurements over time \cite{ueda2007stochastic,berg1977physics,hu2010physical,hu2011geometry,endres2008accuracy,ten2016fundamental}, they can improve gradient sensing. Our results in Appendix \ref{app:snapshot} give the estimator $\hat{\mathbf{g}}$ for the gradient vector $\mathbf{g}$ given the snapshot data. Defining a time-integrated estimator $\hat{\mathbf{g}}_T = \frac{1}{T}\int_0^T \hat{\mathbf{g}}(t) dt$, as in earlier work \cite{hu2010physical,hu2011geometry,camley2017cell}, we  find its variance, $\sigma_{g,T}^2 = \langle |\mathbf{g}_T - \mathbf{g}|^2 \rangle$. To do this, we must consider the kinetics of binding and unbinding at each receptor, which happens with a receptor correlation time of $\tau_i = 1/(k_-^i + c_0 k_+^i)$. In the limit of large averaging times $T \gg \tau_i$, we find (Appendix \ref{app:naive}):

\begin{equation}
    \sigma_{g,T}^2 = \frac{32 \sum_i^R f_i\beta_i \tau_i}{N T \left(\sum_i^R f_i\beta_i\right)^2} \; \; \; \textrm{(naive average)}
\label{eq:variance_naive}
\end{equation}
where $\beta_i = c_0 K_D^i/(c_0+K_D^i)^2$ reflects the accuracy of measuring by type $i$ -- maximized at $c_0 = K_D^i$. For a single type, error is reduced by the effective number of measurements $T/2\tau$, $\sigma_{g,T}^2 = \sigma_g^2 \times 2 \tau/T$ \cite{hu2010physical,hu2011geometry}. \eq \ref{eq:variance_naive} can be cast in a similar form as
\[
\sigma_{g,T}^2 = \sigma_g^2 \sum_i^R \alpha_i \frac{2 \tau_i}{T}
\]
where $\alpha_i = \beta_i f_i / \sum_i^R \beta_i f_i$ is the weight given to receptor type $i$. This equation shows that the variance -- in a naive time average -- is reduced by a weighted sum that depends on the receptor correlation times. In the single receptor type case, $\sigma_{g,T}^2$ becomes arbitrarily small as $\tau \to 0$. However, because $\sigma_{g,T}^2$ is proportional to a weighted {\it sum} of $\tau_i / T$, when receptor correlation times $\tau_i$ decrease, error is limited by the slowest correlation time. If one receptor correlation time is significantly faster than another, \eq \ref{eq:variance_naive} predicts reduced error merely by removing the slow receptors (\fig \ref{fig:SNR}). The naive time average, therefore, does not efficiently use the information available -- if it did, the cell would not be able to gain accuracy by throwing away measurements. The core reason for the failure of naive time averaging is that the snapshot estimator weights receptors equally -- which is appropriate to the amount of information they provide {\it at that moment}. Naively averaging this estimator weighs information from fast receptors (which gain more information as $T$ increases) and slow receptors (which gain less information) similarly. 

The failure of naive time averaging is reminiscent of a well-known result for concentration sensing: it is not optimal for a single receptor to estimate $c$ from a simple time-average of its occupation, $T^{-1}\int_0^T n(t)dt$. If the whole history of binding and unbinding events is used in a maximum likelihood estimate, the error $\sigma_c^2$ is reduced by two \cite{endres2009maximum}. We compute the accuracy limit for gradient sensing $\sigma_{g,T}^2$ using the entire receptor trajectory (Appendix \ref{app:ert}), finding (again in the limit $T \gg \tau_i)$:
\begin{equation}
    \sigma_{g,T;ERT}^2 = \frac{16}{N \sum^R_i f_i \beta_i \frac{T}{\tau_i}} \; \; {\textrm{(entire receptor trajectory)}}
\label{eq:variance_ERT}
\end{equation}
or more intuitively,
\[
    \sigma_{g,T;ERT}^2 = \sigma_g^2 \frac{1}{\sum^R_i \alpha_i \frac{T}{\tau_i}}
\]
For a single receptor type, \eq \ref{eq:variance_ERT} is a factor of two smaller than the naive time average \eq \ref{eq:variance_naive}, precisely as in concentration sensing. However, for multiple types, ERT error can be orders of magnitude better, as the time correlation factors $\tau_i/T$ add ``in parallel'' -- error is no longer limited by the slowest type.

We illustrate the differences between these two errors in \fig \ref{fig:SNR}, computing SNR $g^2/\sigma_g^2$ for two receptor types. Which type provides more information depends on the relative off rates of the two types:
\begin{align}
\textrm{SNR}_{ERT} &= \frac{N g^2 k_-^A T}{16} \left[ f_A \frac{c_0}{c_0 + K_A} + (1-f_A) \rho \frac{c_0}{c_0 + K_B} \right] \label{eq:snr2}
\end{align}
where $\rho = k_-^B/k_-^A$. 

\fig \ref{fig:SNR} shows that as $\rho$ is varied, the naive time average SNR is always {\it at least} a factor of two lower than ERT. When the $B$ receptor off rate is large ($\rho \gg 1$), the naive average is worse than if only the $N/2$ $B$ receptors were used (the ``Naive (fast only)'' yellow dotted line). In this limit, most information is from the $B$ receptors, and using only $B$ receptors reaches half the ERT SNR (\fig \ref{fig:SNR}). 
\begin{figure*}[htbb]
    \centering  
    \includegraphics[width=\linewidth]{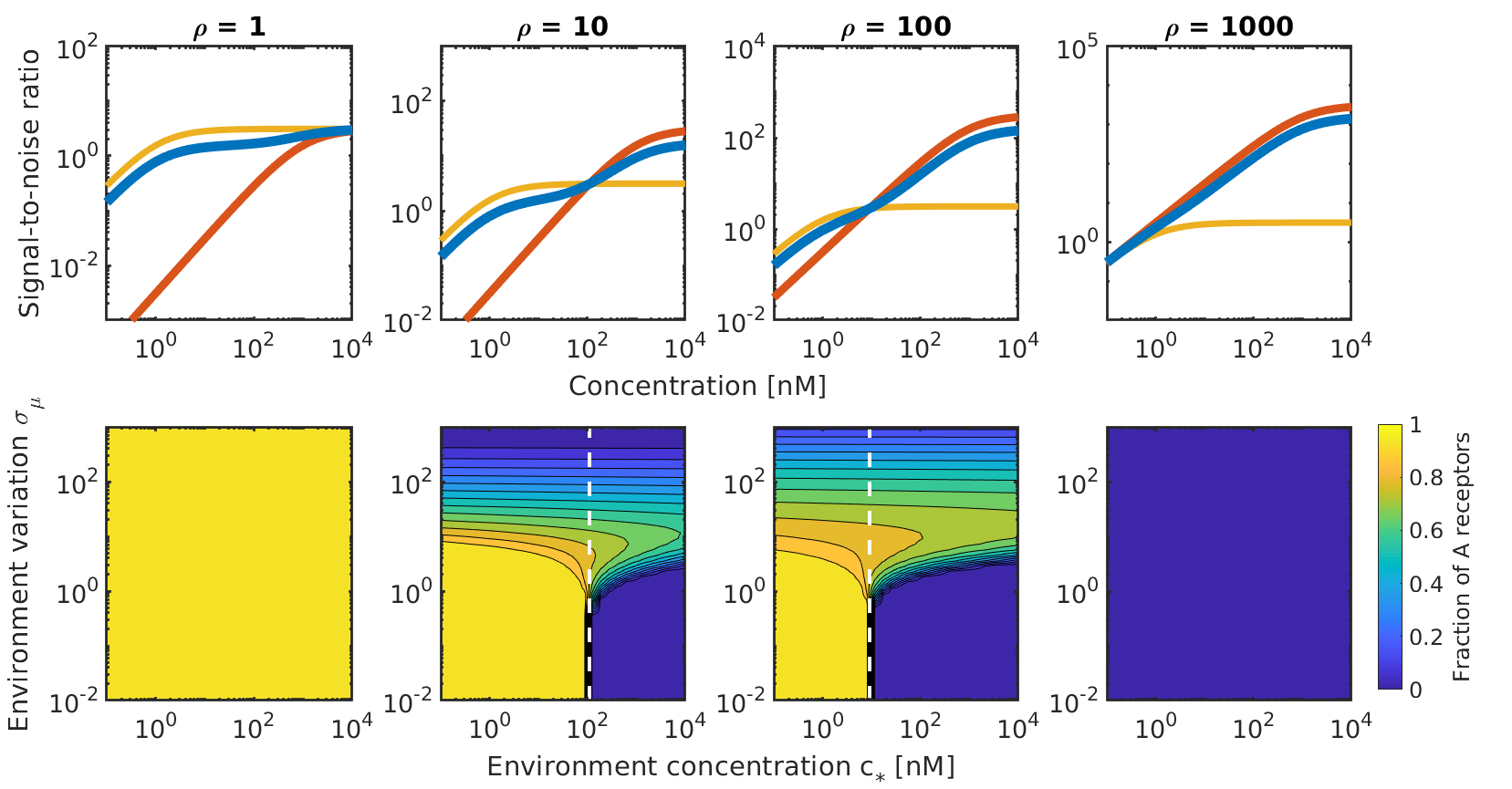}
    \caption{\textbf{Tradeoff between two receptor types depends on their dynamic properties---}. TOP: SNR calculated for differing values of $\rho$ at different fixed concentrations $c_0$ using \eq \ref{eq:snr2}. As in \fig \ref{fig:hedge1}, yellow indicates all-$A$, red all-$B$ and blue a 50-50 mix. BOTTOM: Optimal receptor fraction with time-averaging. Dashed white lines show the analytical $c_{bal}$. Parameters are as in \fig \ref{fig:hedge1} except $N = 10^4$ (chosen as cells have larger SNR with fewer receptors when time-averaging) and $k_-^A T = 2$.}
    \label{fig:timeaverage}
\end{figure*}

How does time averaging affect bet-hedging? For a single environmental concentration, all-$A$ is optimal when \eq \ref{eq:snr2} increases with increasing $f_A$, i.e. $c_0 / (c_0 + K_A) > \rho c_0 / (c_0+K_B)$, or $c_0 < c_{bal} \equiv \frac{K_B - K_A \rho}{\rho - 1}$. \resp{For $\rho \le 1$, it is always best to use the lower-$K_D$ receptor -- but  trade-offs are more complex when $B$ receptors are faster.} The balancing point $c_{bal}$ varies from $c_{bal} \to \infty$ at $\rho = 1$ to $c_{bal} = 0$ at $\rho = K_B/K_A$. $\rho = K_B/K_A$ corresponds to the condition where the on rates of the two types are equal, $k_+^A = k_+^B$, as $k_-^B / k_-^A = K_B/K_A = \frac{k_-^B}{k_-^A}\frac{k_+^A}{k_+^B}$ -- this would be the case if the on rates were diffusion-limited. 

Dependence on receptor off rates is preserved when we study optimal receptor configurations in an uncertain environment $p(\ln c_0) \sim e^{-(\ln c_0 - \ln c_*)^2/2 \sigma_\mu^2}$. In \fig \ref{fig:timeaverage}, we show how the optimal share of $A$ receptors depends on $\rho = k_-^B/k_-^A$. Even with significant uncertainty, when $\rho = 1$, all-$A$ is optimal. However, at higher values of $\rho$, a transition between all-$A$ and all-$B$ like \fig \ref{fig:hedge1} occurs when $c_* = c_{bal}$ (white dashed line). {By contrast with the snapshot results, in the time average at large uncertainties $\sigma_\mu > 1$, the 50-50 mixture is not optimal. Instead, at high uncertainties, for $\rho = 10, 100$}, optimal receptor fractions are similar above and below $c_{bal}$, and the fraction of $A$ receptors exceeds $0.5$, with $f_A$ decreasing as $\sigma_\mu \gtrsim 10$. Why? When cells time-average, CI is nonzero over a broad range of $c_0$ (see \fig \ref{fig:timeaverage}, top), and our argument from the snapshot case fails. 

Can we produce a plot such as \fig \ref{fig:branch} showing the optimal receptor configuration if time-averaging is performed? No. When time averaging, error is minimized by making the correlation times as small as possible -- taking $k_- \to \infty$. We cannot find a consistent set of optimal receptors unless $k_-$ is restricted by some biochemical constraint. This is because, as recognized for concentration sensing \cite{endres2009maximum}, only binding rates are sensitive to $c$ -- bound times should be minimized.

{\it Discussion.---} Our results show that cells can hedge their bets against an uncertain environment by expressing multiple receptor types -- but that this behavior is only reasonable if the uncertainty in $c_0$ spans the range of observed $K_D$ (\fig \ref{fig:branch}) -- i.e. if cells typically explore environments where concentration varies over orders of magnitude. \resp{The idea that signal-processing should be adapted to the likely range of concentrations is similar to classical results showing information transmission is maximized by tuning input-output relationships to the input probability distribution \cite{bialek2012biophysics,barlow1961possible,tkavcik2008information,laughlin1981simple}.} When cells chemotax to hunt bacteria, {as Dictyostelium uses folic acid chemotaxis,} it is intuitively plausible that observed $c_0$ span orders of magnitude, as bacterial hunting must function over both sparse and concentrated solutions, and over many distances to bacteria. {However, at any fixed concentration, expressing multiple receptor types is always suboptimal to choosing the receptor that best fits your current concentration $c_0$ -- so hedging is plausible in circumstances when the environment is uncertain {\it over the timescale on which the receptor affinity is fixed}. What other timescales could appear in the problem? Dictyostelium receptors are internalized in response to large increases in cAMP concentration \cite{serge2011quantification,wang1988localization}, but this process takes several minutes -- much longer than it would typically take Dicty to chemotax to a new mean concentration level. We also show that even if receptor numbers change in different environments, we see very similar results (Appendix \ref{app:number}).} 

\resp{Our work has not so far distinguished between truly different receptors and receptors that are phosphorylated or otherwise modified to change their $K_D$  \cite{islam2018camp,xiao1999desensitization}, which play a role in adaptation to different signal levels in bacteria \cite{tu2018adaptation}. If receptor modification is fast compared to the environment's change in concentration $c_0$, i.e. can occur before the cell samples a new concentration from $P(c_0)$, hedging  will be less effective.
Our work thus suggests interesting future directions for extension, based on recent studies of concentration sensing in time-varying environments \cite{mora2019physical}. } Future work could also consider multiple ligand types, which could also limit accuracy by competing for receptors \cite{mora2015physical,singh2020universal}. 
\resp{For cells to reach the lower bound of \eq \ref{eq:variance_ERT}, they must compute estimates with some reaction network, possibly extending recent work showing how to compute the ERT estimate in concentration sensing \cite{singh2020universal,lang2014thermodynamics}. One important difference here is that finding the ERT $\hat{g}$ requires spatially resolved measurements of bound and unbound times separately for each receptor type. Computation of the ERT estimate for concentration requires additional free energy expenditure \cite{lang2014thermodynamics} -- it would be interesting to determine if the extravagant benefits of the ERT approach for gradient sensing with multiple receptor types (\fig \ref{fig:SNR}) comes with a commensurate cost. Though these are significant complexities, the huge gap between the fundamental bound of \eq \ref{eq:variance_ERT} and the naive average of \eq \ref{eq:variance_naive} shows even very rough approximations to the ERT  provide significant gains over a naive average.}

\begin{acknowledgments}
We thank Wouter-Jan Rappel and Emiliano Perez Ipi\~{n}a for a close reading of the paper. We thank Allyson Sgro for useful discussions. BAC acknowledges support from the grant PHY-1915491.
\end{acknowledgments}

\onecolumngrid
\appendix

\section*{Notation For Appendix}

This appendix includes calculations where we have to distinguish the type $i$ and the index $n$ of each receptor. The intermediate calculations are more complex than the final results in the main paper, and to help keep the details straight, we will denote receptor types $i$ with superscripts and receptor indices $n$ with subscripts. A superscript to a power $i$ like $C_n^i$ never denotes exponentiation. To keep consistent with this, we've included formulas with $\beta^i$ where in the main text we've written $\beta_i$.

\section{Numerical details}
\label{app:numerical}

To evaluate the integral $\cim = \int d c_0 p(c_0) \textrm{CI}(c_0)$ numerically, we found some difficulties that arise because of how broad $p(c_0)$ is. In particular, because the SNR most naturally varies on the log scale (see Fig. \ref{fig:hedge1}, top), it is easiest for us to evaluate this expectation in terms of $\ln c_0$,
\begin{equation}
    \cim = \int d \ln c_0 p(\ln c_0) \textrm{CI}(e^{\ln c_0}) \label{eq:log_expectation}
\end{equation}
For our log-normal distribution, $p(\ln c_0)$ has the simple Gaussian form 
\begin{equation}
    p(\ln c_0)  = \frac{1}{\sigma_\mu \sqrt{2 \pi}} \exp\left[ -\frac{(\ln c_0 - \ln c_*)^2}{2\sigma_\mu^2} \right]
\end{equation}
However, we note that even \eq \ref{eq:log_expectation} can be tricky to evaluate when $\textrm{CI}$ is nonzero only for a range of $\ln c_0$ compared with the scale $\sigma_\mu$. We evaluate this integral using Matlab's Gauss-Kronrod quadrature ({\texttt{quadgk}}), setting waypoints to ensure that all nonzero ranges of both functions should be included. 

To find the receptor type fractions that optimize \cim we use Matlab's \texttt{fminbnd} (golden section search) and \texttt{fminsearch} (Nelder-Mead), depending on the number of variables to be optimized over. 

Code to reproduce the results in the paper can be found at:
\href{https://github.com/bcamley/hedging_reproduce}{https://github.com/bcamley/hedging\_reproduce}

\section{Snapshot sensing}
\label{app:snapshot}

\begin{figure}[htb]
    \centering
    \includegraphics[width=0.4\linewidth]{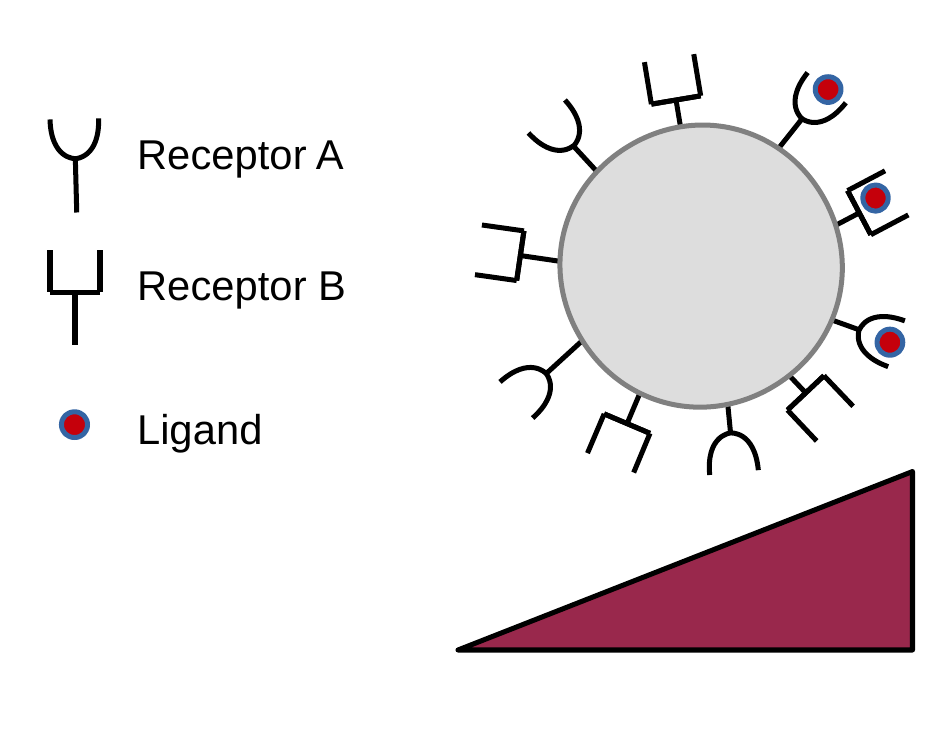}
    \caption{Illustration of data used for gradient estimate using a snapshot of receptor state; only two receptor types are illustrated.}
    \label{fig:snapshot_illustration}
\end{figure}

How precisely can a cell make a measurement of a chemical gradient -- using only its current information about which receptors on its surface are occupied? We extend results from \cite{hu2010physical,hu2011geometry} on this accuracy to include multiple receptor types.
We assume that the cell is in a shallow exponential gradient with direction $\phi$ and {steepness $g=\frac{L|\nabla C|}{C_0}$}, where $L$ is the diameter of the cell {and $C_0$ the concentration at the cell center}.
The gradient $\textbf{g}$ can also be written as $\textbf{g}=(g_x,g_y)=\left(g\textrm{cos}(\phi),g\textrm{sin}(\phi)\right)$.
The concentration at a cell receptor with angular coordinate $\varphi$ can then be written as
\begin{equation}
    C(\varphi) = C_0\textrm{exp}\left[\frac{g}{2}\textrm{cos}(\varphi-\phi)\right].
\label{eq:c_phi}
\end{equation}

Let there be $R$ receptor types, where there are $N^i$ receptors of type $i$ and $N = \sum_i^R N^i$ total receptors. We describe the receptors as being uniformly spread across the cell, with angular positions $\varphi_n^i$ (\fig \ref{fig:snapshot_illustration}).
Then each receptor of type $i$ can be represented as a Bernoulli trial, {i.e. we define a variable  $x_n^i$ that is one if a receptor $n$ of type $i$ is occupied, and zero otherwise. The probability of $x_n^i = 1$ is the probability of that receptor being occupied,}
\begin{equation}
    P_n^i = C_n^i/(C_n^i+K_D^i),
\label{eq:p_occupy}
\end{equation} where $C_n^i$ is the concentration at the $n$th receptor of type $i$ and $K_D^i$ is the dissociation constant of receptor type $i$. 
The dissociation constant $K_D^i$ is the ratio of unbinding and binding rates of the receptors of type $i$, i.e. $K_D^i = \frac{k_{-}^i}{k_{+}^i}$
Assuming that all the receptors are independent of one another, we have the following likelihood function giving the probability of seeing receptor occupations $x_n^i$ given gradient $\mathbf{g}$
\begin{equation}
\nonumber
    \mathcal{L}({\mathbf{g}} | x_1^0,...,x_{N^R}^R) = \prod_i^R \prod_n^{N^i} \left[ \left(P_n^i \right)^{x_n^i} \left( 1- P_n^i \right)^{1-x_n^i} \right].
\end{equation}
The log-likelihood function is
\begin{align}
    \nonumber
    \begin{split}
    &\textrm{ln}\mathcal{L}  = \\ 
    &\sum_i^R \sum_n^{N^i} \left[  x_n^i \textrm{ln}\left( \frac{C_n^i}{C_n^i+K_D^i} \right) + (1-x_n^i) \textrm{ln} \left( \frac{K_D^i}{C_n^i + K_D^i} \right) \right] 
    \end{split}\\
    & = \sum_i^R \sum_n^{N^i} \left[ x_n^i \textrm{ln}\left(\frac{C_n^i}{K_D^i}\right) + \textrm{ln}\left(\frac{K_D^i}{C_n^i+K_D^i}\right) \right] 
\end{align}
In the second term in this equation, we then assume that the receptors are numerous enough that we can replace the sum over receptor position by a continuous integral, $\sum_n^{N^i} \to \frac{N^i}{2\pi}\int_0^{2\pi} d\varphi$:
\begin{align}
    \textrm{ln}\mathcal{L}  = \sum_i^R \left\{ \sum_n^{N^i} \left[ \frac{1}{2}x_n^i g \textrm{cos}\left(\varphi_n^i - \phi\right) + \textrm{ln}\left(\frac{C_0}{K_D}\right)x_n^i \right]  + \int\limits_0^{2\pi} \frac{N^i}{2\pi}\textrm{ln}\left( \frac{K_D^i}{C_0 \textrm{exp}\left(\frac{g}{2}\textrm{cos}\left(\varphi-\phi\right)\right)+K_D^i}\right)d\varphi  \right\}.
\end{align}
We define $z_1^i = \sum\limits_n^{N^i} x_n^i \textrm{cos}\varphi_n^i$ and $z_2^i = \sum\limits_n^{N^i} x_n^i \textrm{sin}\varphi_n^i$, which measure the spatial asymmetry in the occupancy of receptors of type $i$. 
$Z_1 = \sum\limits_i z_1^i$ and $Z_2 = \sum\limits_i z_2^i$ measure the total spatial asymmetry in receptor occupancy of the cell.
In a shallow exponential gradient, we can neglect terms $O(g^4)$ and higher for an estimation of the gradient $\vec{g} = \left( g_x, g_y \right)$.
Then, the log-likelihood function becomes
\begin{equation}
    \textrm{ln} \mathcal{L} = \sum_i^R \left[ \frac{g_x z_1^i + g_y z_2^i}{2} + \textrm{ln} \frac{C_0}{K_D^i} \sum_n^{N^i} x_n^i - \frac{N C_0 f^i K_D^i \left( g_x^2 + g_y^2 \right)}{16(C_0+K_D^i)^2} {+\ln \frac{C_0}{C_0+K_D^i}} \right].
\label{eq:loglike}
\end{equation}
(We note that past papers with similar derivations \cite{hu2011geometry,hopkins2019leader} have not always written the last term in this log-likelihood, which is an irrelevant constant.) Taking the derivative with respect to $g_{x,y}$ gives
\begin{equation}
    \frac{\partial}{\partial g_{x,y}} \textrm{ln} \mathcal{L} = \sum_i^R \left[ \frac{z_{1,2}^i}{2} - \frac{N C_0 f^i K_D^i g_{x,y}}{8(C_0+K_D^i)^2} \right]
    \label{eq:dloglikedalpha}
\end{equation}
Because the log function is monotonic, we can set \eq \ref{eq:dloglikedalpha} equal to zero to find $\hat{g}_x$ and $\hat{g}_y$, the parameters which for $g_x$ and $g_y$ which maximize the likelihood function.
Carrying out this procedure, we find:
\begin{align}
    & \sum_i^R \left[ \frac{z_{1,2}^i}{2} - \frac{N C_0 f^i K_D^i }{8(C_0+K_D^i)^2}\hat{g}_{x,y} \right] = 0 \\
    & \sum_i^R \frac{z_{1,2}^i}{2} = \sum_i^R \left[  \frac{N C_0 f^i K_D^i }{8(C_0+K_D^i)^2}\hat{g}_{x,y} \right] \\
    & \frac{Z_{1,2}}{2} = \frac{N C_0}{8} \sum_i^R \left[ \frac{f^i K_D^i}{(C_0 +K_D^i)^2}\hat{g}_{x,y} \right],
\end{align}
which be solved for $\hat{g}_{x,y}$ to determine estimators ${\hat{g}_{x,y}}$ as
\begin{equation}
    \hat{g}_x = \frac{4 Z_1}{N C_0 \sum\limits_i^R \left[\frac{f^i K_D^i}{(C_0 + K_D^i)^2}\right]} \textrm{ and } \hat{g}_y = \frac{4 Z_2}{N C_0 \sum\limits_i^R \left[\frac{f^i K_D^i}{(C_0 + K_D^i)^2}\right]}.
    \label{eq:alpha_ests}
\end{equation}
To determine the asymptotic variance on these estimators, we will need to compute the second derivative of the log-likelihood function.
Applying an additional derivative to \eq \ref{eq:dloglikedalpha} gives
\begin{equation}
\begin{split}
    \frac{\partial^2}{\partial g_{x,y}^2} \textrm{ln} \mathcal{L} &= - \frac{N C_0}{8} \sum_i^R \left[ \frac{f^i K_D^i}{(C_0+K_D^i)^2} \right] \\  \frac{\partial^2}{\partial g_x \partial g_y} \textrm{ln} \mathcal{L} &= 0
    \label{eq:d2loglikedalpha}
\end{split}
\end{equation}
From the log-likelihood function, we can also determine the Fisher information matrix, {which controls the best possible measurement that the cell can make of the uncertain $\mathbf{g}$ \cite{hu2010physical,kay1993fundamentals}}.
In this case it is diagonal, and its inverse gives the variances of $\hat{g}_x$ and $\hat{g}_y$ in the limit of many samples.
As a result, we have expressions for the asymptotic variances for $\hat{g}_x$ and $\hat{g}_y$
\begin{align*}
    \frac{1}{\sigma_{g_{x,y}}^2} = \left\langle \left(\frac{\partial \textrm{ln} \mathcal{L}}{\partial g_{x,y}}\right)^2 \right\rangle &= -\left\langle \frac{\partial^2 \textrm{ln} \mathcal{L}}{\partial g_{x,y}^2} \right\rangle \\
    &= \frac{N C_0}{8} \sum_i^R \left[ \frac{f^i K_D^i}{(C_0+K_D^i)^2} \right] .
\end{align*}
and so
\begin{equation}
\sigma_{g_{x,y}}^2 = \frac{8}{N C_0 \sum\limits_i^R \frac{f^i K_D^i}{(C_0 + K_D^i)^2}}.    
\end{equation}
The important parameter is $\sigma^2_{\textbf{g}},$ which is just the sum of the component variances
\begin{equation}
    \sigma_{\textbf{g}}^2 =\sigma_{g_x}^2 + \sigma_{g_x}^2 = \frac{16}{N C_0 \sum\limits_i^R \frac{f^i K_D^i}{(C_0 + K_D^i)^2}}.
\label{eq:sig2_g}
\end{equation}
As the sample size becomes large, the distribution of $\hat{g}_{1,2}$ converges to a normal distribution with means $g_{x,y}$ and variance $\sigma_{{g_x},{g_y}}^2$.
This also implies that the mean values of $Z_1$ and $Z_2$ are
\begin{equation}
    \left\langle Z_{1,2} \right\rangle = \frac{1}{4}N C_0 \sum_i^R  \left[\frac{f^i K_D^i}{(C_0 + K_D^i)^2}\right] g_{x,y}.
\label{eq:Z_mean}
\end{equation}

\section{Naive time averaging}
\label{app:naive}

A cell may improve its estimation of the gradient by time averaging. In the previous section, we determined an estimator $\hat{\mathbf{g}}$ that is the best estimate of a cell's gradient, given a snapshot of its receptor information. Naively, a cell could improve its accuracy by making a measurement over a time $T$ and determining the average of these estimates
\begin{equation}
    \hat{\mathbf{g}}_T = \frac{1}{T}\int_0^T dt \, \, \hat{\mathbf{g}}(t)
\end{equation}
Then the variance of this new estimator will be reduced,
\begin{equation}
\begin{split}
    \sigma_{g,T}^2 &= \langle |\hat{\mathbf{g}}_T|^2 \rangle - \langle \hat{\mathbf{g}}_T \rangle^2 \\&= \frac{1}{T^2}\int_0^Tdt\int_0^Tds \left( \left \langle \hat{\textbf{g}}(s)\hat{\textbf{g}}(t) \right\rangle - \left\langle \hat{\textbf{g}} \right\rangle^2 \right).
    \end{split}
\end{equation}

To understand how time averaging improves the cell's sensing accuracy, we need to compute $\langle \hat{\textbf{g}}(s) \cdot \hat{\textbf{g}}(t) \rangle$, the correlation function of $\hat{\textbf{g}}$.
This correlation function is related to the correlation functions in the estimates of each component of the gradient as
\begin{equation}
    \langle \hat{\textbf{g}}(s) \cdot \hat{\textbf{g}}(t) \rangle = \langle \hat{g}_x(s)\hat{g}_x(t) \rangle + \langle \hat{g}_y(s)\hat{g}_y(t) \rangle.
\end{equation}
And, by \eq \ref{eq:alpha_ests}, the correlation functions for $g_{x,y}$ can be related to the correlation functions for $Z_{1,2}$ as
\begin{equation}
    \langle \hat{g}_{x,y}(s)\hat{g}_{x,y}(t) \rangle = \frac{16 \langle Z_{1,2}(s) Z_{1,2}(t) \rangle}{N^2 C_0^2 \left( \sum\limits_i^R \left[\frac{f^i K_D^i}{(C_0 + K_D^i)^2}\right]\right)^2}  .
\label{eq:Z_corr_to_alpha}
\end{equation}

The correlation functions for $Z_1$ can be written in terms of the single receptor correlation function:
\begin{align}
\langle Z_1(s) Z_1(t) \rangle = & \left\langle \left( \sum_i^R \left[ \sum_n^{N^i} x_n^i(s) \textrm{cos}(\varphi_n^i) \right] \right) \left( \sum_i^R \left[ \sum_n^{N^i} x_n^i(t) \textrm{cos}(\varphi_n^i) \right] \right) \right\rangle \\
= & \left\langle \sum_i^R \sum_j^R \sum_n^{N^i} \sum_m^{N^j} x_n^i(s) \textrm{cos}(\varphi_n^i) x_m^j(t) \textrm{cos}(\varphi_m^j) \right\rangle \\
= &  \sum_i^R \sum_j^R \sum_n^{N^i} \sum_m^{N^j} \left\langle x_n^i(s) x_m^j(t) \right\rangle \textrm{cos}(\varphi_n^i) \textrm{cos}(\varphi_m^j).
\label{eq:Z1_corr_sing}
\end{align}
{The kinetics of receptor binding and unbinding with multiple receptor types can be quite complicated \cite{wang2007quantifying,berezhkovskii2013effect}, with ligands potentially diffusing from one receptor to another. However, if the binding and unbinding process is slow with respect to this diffusion -- i.e. binding is reaction-limited, as is believed to be the case in eukaryotic chemotaxis \cite{hu2011geometry,wang2007quantifying}, it is appropriate to think of the ligand-receptor binding having two states -- one bound and one with ligand in the bulk.  Then, there are two relevant rates, that of ligand binding to a receptor of type $i$ exposed to concentration $C$ is $k_+^i C$ and the off rate is $k_-^i$ -- which results in an exponential single receptor correlation function}
\begin{equation}
    \langle x_n^i(s) x_n^i(t) \rangle = \sigma_{x_n^i}^2 e^{-|t-s|/\tau_n^i} + \langle x_n^i \rangle ^2 
\label{eq:sing_corr}
\end{equation}
for receptor $n$ of type $i$. {This limit is also appropriate if all ligand is internalized, as discussed by \cite{endres2009maximum}.}
The parameter $\sigma_{x_n^i}^2$ characterizes the fluctuations in the occupancy of the receptor, and is given by
\begin{equation}
    \sigma_{x_n^i}^2 = \frac{C_n^i K_D^i}{(C_n^i + K_D^i)^2},
\end{equation}
the variance of a Bernoulli trial.
$\tau^i_n$ is the single receptor correlation time
\begin{equation}
    \tau_n^i = 1/(k_{-}^i + C_n^i k_{+}^i)
\label{eq:recep_tau}
\end{equation}
in the reaction-limited case. (Generalization to other limits is possible but not straightforward \cite{kaizu2014berg,bialek2005physical,berezhkovskii2013effect}.)
Because different receptors are independent, the mean of their product is just the product of their means
\begin{equation}
    \langle x_n^i(s) x_m^j(t) \rangle = \langle x_n^i(s) \rangle \langle x_m^j(t) \rangle \; \; \; \; \textrm{if i$\neq$ j or n$\neq$m}.
\label{eq:ind_receps}
\end{equation}
Using \eq \ref{eq:sing_corr} for terms where $i=j$ and $n=m$ and \eq \ref{eq:ind_receps} otherwise, we can expand the correlation function of $Z_1$ in \eq \ref{eq:Z1_corr_sing} as
\begin{align}
    \langle Z_1(s) Z_1(t) \rangle = & \sum_i^R \sum_j^R \sum_n^{N^i} \sum_m^{N^j} \left\langle x_n^i(s) x_m^j(t) \right\rangle \textrm{cos}(\varphi_n^i) \textrm{cos}(\varphi_m^j) \\
    = & \sum_i^R\sum_n^{N^i} \sigma_{x_n^i}^2 e^{-|t-s|/\tau_n^i} \textrm{cos}^2(\varphi_n^i) + \sum_i^R \sum_j^R \sum_n^{N^i} \sum_m^{N^j} \langle x_n^i(s) \rangle \langle x_m^j(t) \rangle \textrm{cos}(\varphi_n^i) \textrm{cos}(\varphi_m^j).
    \label{eq:Z1_corr_expand}
\end{align}
The second term in \eq \ref{eq:Z1_corr_expand} is $\langle Z_1 \rangle^2$, which can be solved as
\begin{equation}
    \langle Z_1 \rangle^2 = \frac{1}{16}N^2 C_0^2 \left(\sum_i^R  \left[\frac{f^i K_D^i}{(C_0 + K_D^i)^2}\right]\right)^2 g_x^2.
\end{equation}
by \eq \ref{eq:Z_mean}.
For the first term in \eq \ref{eq:Z1_corr_expand}, taking the sum to an integral gives
\begin{equation}
\sum_i^R\sum_n^{N^i} \sigma_{x_n^i}^2 e^{-|t-s|/\tau_n^i}\textrm{cos}^2(\varphi_n^i) = \frac{N C_0}{2} \sum_i^R f^i \frac{K_D^i}{\left(C_0+K_D^i\right)^2}e^{-|t-s|/\tau^i} + O(g^2),
\end{equation}
where $\tau^i = 1/(k_{-}^i + C_0 k_{+}^i)$, i.e., \eq \ref{eq:recep_tau} for a receptor in the ambient concentration $C_0$.
Therefore, for shallow gradients, the correlation function for $Z_1$ is
\begin{equation}
    \langle Z_1(s) Z_1(t) \rangle = \frac{N C_0}{2} \sum_i^R f^i \frac{K_D^i}{\left(C_0+K_D^i\right)^2}e^{-|t-s|/\tau^i} + \frac{1}{16}N^2 C_0^2 \left(\sum_i^R  \left[\frac{f^i K_D^i}{(C_0 + K_D^i)^2}\right]\right)^2 g_x^2.
\label{eq:Z1_corr}
\end{equation}
Using the relation in \eq \ref{eq:Z_corr_to_alpha}, the correlation function of the estimator $\hat{g}_1$ can be found from \eq \ref{eq:Z1_corr}:
\begin{equation}
    \langle \hat{g}_1(s) \hat{g}_1(t) \rangle = \frac{8 \sum_i^R \left[ \frac{f^i K_D^i}{(C_0 + K_D^i)^2} e^{-|t-s|/\tau^i} \right]}{N C_0 \left( \sum_i^R \frac{f^i K_D^i}{(C_0 + K_D^i)^2}\right)^2} + g_x^2.
\label{eq:alpha_1_corr}
\end{equation}
Similar expressions for the correlation functions of $Z_2$ and $\hat{g}_2$ can be derived as
\begin{equation}
    \langle Z_2(s) Z_2(t) \rangle = \frac{N C_0}{2}\sum_i^R f^i \frac{K_D^i}{\left(C_0+K_D^i\right)^2}e^{-|t-s|/\tau^i} + \frac{1}{16}N^2 C_0^2 \left(\sum_i^R  \left[\frac{f^i K_D^i}{(C_0 + K_D^i)^2}\right]\right)^2 g_y^2
\label{eq:Z2_corr}
\end{equation}
and
\begin{equation}
    \langle \hat{g}_2(s) \hat{g}_2(t) \rangle = \frac{8 \sum_i^R \left[ \frac{f^i K_D^i}{(C_0 + K_D^i)^2} e^{-|t-s|/\tau^i} \right]}{N C_0 \left( \sum_i^R \frac{f^i K_D^i}{(C_0 + K_D^i)^2}\right)^2} + g_y^2.
\label{eq:alpha_2_corr}
\end{equation}
Thus, the correlation in $\hat{
\textbf{g}}$ for shallow gradients is 
\begin{equation}
    \langle \hat{\textbf{g}}(s) \cdot \hat{\textbf{g}}(t) \rangle = \frac{16 \sum_i^R \left[ \frac{f^i K_D^i}{(C_0 + K_D^i)^2} e^{-|t-s|/\tau^i} \right]}{N C_0 \left( \sum_i^R \frac{f^i K_D^i}{(C_0 + K_D^i)^2}\right)^2} + g^2.
\label{eq:gcorr}
\end{equation}

This now gives us enough information to compute the time-averaged variance,
\begin{equation}
    \sigma_{g,T}^2 = \frac{1}{T^2}\int_0^Tdt\int_0^Tds \left( \left \langle \hat{\textbf{g}}(s)\hat{\textbf{g}}(t) \right\rangle - \left\langle \hat{\textbf{g}} \right\rangle^2 \right).
\end{equation}
Then, using the result in \eq \ref{eq:gcorr}, we have
\begin{align}
    \sigma_{g,T}^2 & =  \frac{16 \sum_i^R \left[ \frac{f^i K_D^i}{(C_0 + K_D^i)^2} \int_0^Tdt\int_0^Tds e^{-|t-s|/\tau^i} \right]}{T^2 N C_0 \left( \sum_i^R \frac{f^i K_D^i}{(C_0 + K_D^i)^2}\right)^2} \\
    & = 2 \sigma_{\textbf{g}}^2 \frac{\sum_i^R \frac{f^i K_D^i}{(C_0 + K_D^i)^2} \tau^i\left[T-\tau^i\left(1-e^{-T/\tau^i}\right)\right]}{T^2 \sum_i^R \frac{f^i K_D^i}{(C_0 + K_D^i)^2}}.
\label{eq:simple_average}
\end{align}
In the limit where $T \gg \tau^i$ for all $\tau^i$, \eq \ref{eq:simple_average} becomes
\begin{equation}
    \sigma_{g,T}^2 = 2\sigma_{\textbf{g}}^2 \frac{\sum_i^R \frac{f^i K_D^i}{(C_0 + K_D^i)^2} \tau^i}{T \sum_i^R \frac{f^i K_D^i}{(C_0 + K_D^i)^2}} = \frac{32 \sum_i^R f^i\beta^i \tau^i}{N T \left(\sum_i^R f^i\beta^i\right)^2},
\label{eq:sig2g_avg}
\end{equation}
where the parameter $\beta^i = C_0 K_D^i/(C_0+K_D^i)^2$ reflects the accuracy of measuring only using receptor $i$, as in the main text.

\begin{figure}[htb]
    \centering
    \includegraphics[width=\linewidth]{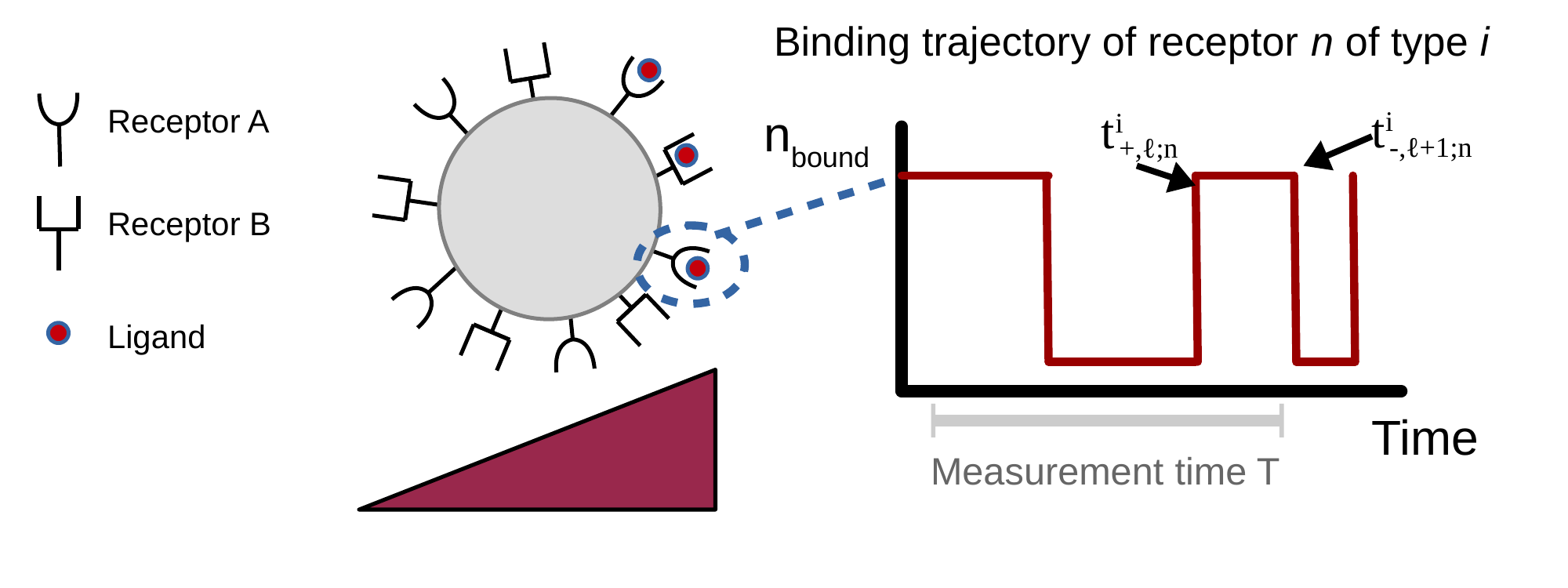}
    \caption{Illustration of data used for maximum likelihood estimate from entire receptor trajectories (ERT)}
    \label{fig:mle_illustration}
\end{figure}

\section{Maximum likelihood using entire receptor binding trajectory (ERT)}
\label{app:ert}
Instead of simply performing a naive average, a cell could also improve its sensing of the gradient by determining an estimate of the gradient from the history of its receptors over the measurement time -- when they are bound and unbound (\fig \ref{fig:mle_illustration}).
We a time interval for receptor $n$ of type $i$ as time series $\{t^i_{+;n},t^i_{-;n}\}$, where particles bind at times $t^i_{+;n,\ell}$ and unbind at times $t^i_{-;n,\ell}$, where $\ell$ indexes the binding and unbinding events.
Following \cite{endres2009maximum}, we compute the probability for a time series of binding and unbinding events.
Define a function
\begin{equation}\nonumber 
    f^{*;i}_{-;n}(t^i_{-,\ell;n}) = p^i_{-,\ell;n}(t^i_{-,\ell;n}|t^i_{+,1;n},t^i_{-,1;n}, \dots ,t^i_{-,\ell-1;n},t^i_{+,\ell;n}) 
\end{equation}
which is the probability density for the event that receptor $n$ of type $i$ experiences an unbinding at time $t^i_{-,\ell;n}$ given the previous time series data $\{t^i_{+,1;n},t^i_{-,1;n}, \dots ,t^i_{-,\ell-1;n},t^i_{+,\ell;n}\}$.
{Here, the time series has been written for a receptor that is initially unbound, and the indexing of the time series in the following equations will follow that notation. The procedure is the same for a receptor that starts in the bound state.}
Define the analogous function
\begin{equation}
\nonumber
    f^{*;i}_{+;n}(t^i_{+,\ell;n}) =p^i_{+,\ell;n}(t^i_{+,\ell;n}|t^i_{+,1;n},t^i_{-,1;n}, \dots ,t^i_{+,\ell-1;n},t^i_{-,\ell-1;n})
\end{equation}
for binding events.
Then, the probability of observing a time series $\{t^i_{+;n},t^i_{-;n}\}$ is given by
\begin{equation}
    p(\{t^i_{+;n},t^i_{-;n}\}) = \prod_\ell^{\eta^i_{b;n}} f^{*;i}_{-;n}(t^i_{-,\ell;n})\prod_{\ell^\prime}^{\eta^i_{u;n}}f^{*;i}_{+;n}(t^i_{+,{\ell^\prime};n})
\label{eq:one_timeseries}
\end{equation}
where $\eta^i_{b;n},\eta^i_{u;n}$ are the numbers of binding events and unbinding events, respectively.
If a cell measures for a time interval $T$ that is long compared to the relevant time scales (ie, $T\gg 1/k_{-}^i$, $T \gg 1/C_0k_{+}^i$ for all receptor types), then $\eta^i_{b;n} \approx \eta^i_{u;n}$ because the number of binding events can differ by at most one from the number of unbinding events.
In this limit, the information about the gradient is dominated by the observed time series, and not the initial snapshot state of the receptors. 

We assume that we are in the reaction-limited case, where we can treat the rate of binding to a receptor of type $i$ exposed to concentration $C$ as $k_+^i C$ and the off rate as $k_-^i$ -- neglecting rebinding. {(Neglecting rebinding, as discussed in more detail by \cite{endres2009maximum}, is also the appropriate limit to find the fundamental bound to accuracy, as cells may prohibit rebinding by degrading or internalizing ligand.)}
The functions $f^{*;i}_{+;n}(t^i_{-,\ell;n})$ and $f^{*;i}_{-;n}(t^i_{-,\ell;n})$, with this simple Markovian kinetics, do not depend on the whole time series, they only depend on the time of previous unbinding/binding event:
\begin{align}
    & f^{*;i}_{-;n}(t^i_{-,\ell;n}) = k^i_{-}e^{-k^i_{-}(t^i_{-,\ell;n}-t^i_{+,\ell-1;n})}\\
    & f^{*;i}_{+;n}(t^i_{+,\ell;n}) = k^i_{+}C^i_n e^{-k^i_{+}C^i_n(t^i_{+,\ell;n}-t^i_{-,\ell;n})}.
\end{align}
These are probability density functions for exponential distributions with rates $k^i_{-}$ and $k^i_{+}C^i_n$ for unbinding and binding, respectively.
From these equations, we can determine the likelihood function for the gradient parameters $g_x$ and $g_y$ given the observed time series $\{t^i_{+;n},t^i_{-;n}\}$ at receptors $n$ of types $i$.
Because each receptor is independent, the likelihood function is the product of the probability function in \eq \ref{eq:one_timeseries}:
\begin{equation}
    \mathcal{L} = \prod^R_i \prod^{N^i}_n \left( \prod_\ell^{\eta^i_{b;n}} k^i_{-}e^{-k^i_{-}(t^i_{-,\ell;n}-t^i_{+,\ell-1;n})}\prod_{\ell^\prime}^{\eta^i_{u;n}}k^i_{+}C^i_n e^{-k^i_{+}C^i_n(t^i_{+,\ell;n}-t^i_{-,{\ell^\prime};n})} \right).
\label{eq:likelihood_timeseries}
\end{equation}
Then, define the total time bound $\mathcal{T}^i_{b;n}$ and the total time unbound $\mathcal{T}^i_{u;n}$ for receptor $n$ of type $i$
{(the indexing in these definitions assume the receptor starts unbound, but analogous definitions can be written for a receptor that is bound at $t=0$, and in the limit of long times treated here, this assumption does not matter.)}
\begin{align}
    & \mathcal{T}^i_{b;n} = \sum^{\eta^i_{b;n}}_\ell t^i_{-,\ell;n}-t^i_{+,\ell;n} \\
    & \mathcal{T}^i_{u;n} = \sum^{\eta^i_{u;n}}_\ell t^i_{+,\ell;n}-t^i_{-,\ell-1;n}.
\end{align}
With these definitions, the likelihood function in \eq \ref{eq:likelihood_timeseries} becomes
\begin{equation} \nonumber
    \mathcal{L} = \prod^R_i \prod^{N^i}_n \left(k^i_{-}\right)^{\eta^i_{b;n}}\left(k^i_{+}C^i_n\right)^{\eta^i_{u;n}}e^{-k^i_{-}\mathcal{T}^i_{b;n}}e^{-k^i_{+}C^i_n\mathcal{T}^i_{u;n}}.
\end{equation}
The log-likelihood function is then
\begin{equation}
    \textrm{ln} \mathcal{L} = \sum^R_i \sum^{N^i}_n \left[ \eta^i_{b;n}\textrm{ln}\left(k^i_{-}\right) + \eta^i_{u;n}\textrm{ln}\left(k^i_{+}C^i_n\right) -k^i_{-}\mathcal{T}^i_{b;n}-k^i_{+}C^i_n\mathcal{T}^i_{u;n} \right].
\label{eq:loglike_ts}
\end{equation}
Substituting the expression $C_n^i = C_0\textrm{exp}\left[\frac{1}{2}\left(g_x\textrm{cos}(\phi^i_n)+g_y\textrm{sin}(\phi^i_n)\right)\right]$ in \eq \ref{eq:loglike_ts} gives
\begin{equation}
\begin{split}
    \textrm{ln} \mathcal{L} =  \sum^R_i \sum^{N^i}_n  \Bigg\{ \eta^i_{b;n}\textrm{ln}\left(k^i_{-}\right) + \eta^i_{u;n}\textrm{ln}\left(k^i_{+}C_0\right) &+ \eta^i_{u;n}\left[\frac{1}{2}\left(g_x\textrm{cos}(\phi^i_n)+g_y\textrm{sin}(\phi^i_n)\right)\right] \\&-k^i_{-}\mathcal{T}^i_{b;n}-k^i_{+}C_0\textrm{exp}\left[\frac{1}{2}\left(g_x\textrm{cos}(\phi^i_n)+g_y\textrm{sin}(\phi^i_n)\right)\right]\mathcal{T}^i_{u;n} \Bigg\}.
\end{split}
\end{equation}
For shallow gradients, we can approximate the log-likelihood function by expanding to second order in the magnitude of the gradient.
This results in
\begin{equation}
\begin{split}
    \textrm{ln} \mathcal{L} \approx  \sum^R_i \sum^{N^i}_n  \Bigg\{ \eta^i_{b;n}\textrm{ln}\left(k^i_{-}\right) &+ \eta^i_{u;n}\textrm{ln}\left(k^i_{+}C_0\right) +  \eta^i_{u;n}\left[\frac{1}{2}\left(g_x\textrm{cos}(\phi^i_n)+g_y\textrm{sin}(\phi^i_n)\right)\right]-k^i_{-}\mathcal{T}^i_{b;n} \\
    &-k^i_{+}C_0\left[1+\frac{1}{2}\left(g_x\textrm{cos}(\phi^i_n)+g_y\textrm{sin}(\phi^i_n)\right)+\frac{1}{8}\left(g_x\textrm{cos}(\phi^i_n)+g_y\textrm{sin}(\phi^i_n)\right)^2\right]\mathcal{T}^i_{u;n} \Bigg\}.
\end{split}
\label{eq:loglike_ts_shallow}
\end{equation}
Differentiating \eq \ref{eq:loglike_ts_shallow} with respect to $g_x$ and $g_y$, we get
\begin{equation}
    \frac{\partial}{\partial g_x} \textrm{ln} \mathcal{L} = \sum^R_i \sum^{N^i}_n \Bigg\{ \frac{\eta^i_{u;n}}{2}\textrm{cos}(\phi^i_n)-k^i_{+}C_0\left[\frac{1}{2}\textrm{cos}(\phi^i_n)+\frac{1}{4}\left(g_x\textrm{cos}(\phi^i_n)+g_y\textrm{sin}(\phi^i_n)\right)\textrm{cos}(\phi^i_n)\right]\mathcal{T}^i_{u;n} \Bigg\}
    \label{eq:dlogliketsda1}
\end{equation}
\begin{equation}
    \frac{\partial}{\partial g_y} \textrm{ln} \mathcal{L} = \sum^R_i \sum^{N^i}_n \Bigg\{ \frac{\eta^i_{u;n}}{2}\textrm{sin}(\phi^i_n)-k^i_{+}C_0\left[\frac{1}{2}\textrm{sin}(\phi^i_n)+\frac{1}{4}\left(g_x\textrm{cos}(\phi^i_n)+g_y\textrm{sin}(\phi^i_n)\right)\textrm{sin}(\phi^i_n)\right]\mathcal{T}^i_{u;n} \Bigg\}.
    \label{eq:dlogliketsda2}
\end{equation}
Because $g_x$ and $g_y$ here do not depend on $i$ and $n$, \eq \ref{eq:dlogliketsda2} can be equated to zero and solved to find the maximum likelihood estimator in terms of sums over  functions of $\mathcal{T}^i_{u;n}$ and $\phi^i_n$. However, we have not found the precise form very useful. 
From the derivatives of the log-likelihood function, we can compute the Fisher Information Matrix:
\begin{equation}
    \mathcal{I}_{a,b} = -\left\langle  \frac{\partial^2}{\partial a \partial b} \textrm{ln} \mathcal{L} \right\rangle.
\end{equation}
Differentiating \eq \ref{eq:dlogliketsda1} and \eq \ref{eq:dlogliketsda2} with respect to combinations of $g_x$ and $g_y$ gives the following matrix
\begin{equation}
    \mathcal{I} = 
    \begin{bmatrix}
    \sum^R_i \sum^{N^i}_n \frac{k^i_{+}C_0\langle \mathcal{T}^i_{u;n} \rangle}{4} \textrm{cos}^2(\phi^i_n) & \sum^R_i \sum^{N^i}_n \frac{k^i_{+}C_0\langle \mathcal{T}^i_{u;n} \rangle}{4} \textrm{cos}(\phi^i_n)\textrm{sin}(\phi^i_n) \\
    \sum^R_i \sum^{N^i}_n \frac{k^i_{+}C_0\langle \mathcal{T}^i_{u;n} \rangle}{4} \textrm{cos}(\phi^i_n)\textrm{sin}(\phi^i_n) & \sum^R_i \sum^{N^i}_n \frac{k^i_{+}C_0\langle \mathcal{T}^i_{u;n} \rangle}{4} \textrm{sin}^2(\phi^i_n) \\
    \end{bmatrix}.
\label{eq:FI_matrix}
\end{equation}
The expectation value $\langle \mathcal{T}^i_{u;n} \rangle$ can be found in terms of the measurement time $T$ and the probability $P^i_n$ that a receptor is occupied (\eq \ref{eq:p_occupy})
\begin{equation}
    \langle \mathcal{T}^i_{u;n} \rangle = T(1-P^i_n) = T\frac{K^i_d}{C^i_n+K^i_d}.
\label{eq:unbound_time}
\end{equation}
By substituting \eq \ref{eq:unbound_time} into \eq \ref{eq:FI_matrix} and taking the inner sums to an integral, we have the following expressions for each matrix element:
\begin{equation}
\begin{split}
     \sum^R_i \sum^{N^i}_n \frac{k^i_{+}C_0\langle \mathcal{T}^i_{u;n} \rangle}{4} \textrm{cos}^2(\phi^i_n) & = \frac{1}{4}\sum^R_i \frac{N^i}{2\pi}\int^{2\pi}_0 \frac{k^i_{+}C_0TK^i_d}{(C_0\textrm{exp}\left[\frac{g}{2}\textrm{cos}(\varphi-\phi)\right] + K^i_d)}\textrm{cos}^2(\phi)d\phi \\
     & = N\sum^R_i \frac{f^ik^i_{-}C_0T}{8(C_0+K^i_d)} + O(g^2)
\end{split}
\end{equation}
\begin{equation}
\begin{split}
     \sum^R_i \sum^{N^i}_n \frac{k^i_{+}C_0\langle \mathcal{T}^i_{u;n} \rangle}{4} \textrm{sin}^2(\phi^i_n) & = \frac{1}{4}\sum^R_i \frac{N^i}{2\pi}\int^{2\pi}_0 \frac{k^i_{+}C_0TK^i_d}{(C_0\textrm{exp}\left[\frac{g}{2}\textrm{cos}(\varphi-\phi)\right] + K^i_d)}\textrm{sin}^2(\phi)d\phi \\
     & = N\sum^R_i \frac{f^ik^i_{-}C_0T}{8(C_0+K^i_d)} + O(g^2)
\end{split}
\end{equation}
\begin{equation}
\begin{split}
     \sum^R_i \sum^{N^i}_n \frac{k^i_{+}C_0\langle \mathcal{T}^i_{u;n} \rangle}{4} \textrm{cos}(\phi^i_n)\textrm{sin}(\phi^i_n) & = \frac{1}{4}\sum^R_i \frac{N^i}{2\pi}\int^{2\pi}_0 \frac{k^i_{+}C_0TK^i_d}{(C_0\textrm{exp}\left[\frac{g}{2}\textrm{cos}(\varphi-\phi)\right] + K^i_d)}\textrm{cos}(\phi^i_n)\textrm{sin}(\phi^i_n)d\phi = 0.
\end{split}
\end{equation}

Therefore, the Fisher information matrix is diagonal, and in shallow gradients it is
\begin{equation}
    \mathcal{I} = 
    \begin{bmatrix}
    NC_0T\sum^R_i \frac{f^ik^i_{-}}{8(C_0+K^i_d)} & 0 \\
    0 & NC_0T\sum^R_i \frac{f^ik^i_{-}}{8(C_0+K^i_d)} \\
    \end{bmatrix}.
\label{eq:FI_matrix_fin}
\end{equation}
We note that \eq \ref{eq:FI_matrix_fin} has only been calculated in the large-$T$ limit; in the limit of $T \to 0$, we would expect the Fisher information to limit to the estimate from a single snapshot. 

For cells with a single receptor type, \eq \ref{eq:FI_matrix_fin} implies that the asymptotic variances on $g_x$ and $g_y$ are 1/2 of their value determined from time averaging---the same factor as in concentration sensing \cite{endres2009maximum}.
However, in the multiple receptor type case, there is a more significant difference.
The variance in $\hat{\textbf{g}}$ determined from \eq \ref{eq:FI_matrix_fin} is the sum of the inverses of the diagonal elements
\begin{equation}
    \begin{split}
    \sigma_{g,T;ERT}^2 &= \frac{16}{NC_0T \sum^R_i \frac{f^ik^i_{-}}{(C_0+K^i_d)}}\\  &= \frac{16}{NC_0 \sum^R_i \frac{f^iK^i_d}{(C_0+K^i_d)^2}\frac{T}{\tau^i}}.
    \end{split}
\label{eq:sig2g_ts}
\end{equation}
As discussed in the main text, this shows that a slow receptor correlation does not act as a limiting factor when the entire receptor trajectory is considered.

\section{Hedging allowing the number of receptors to change}
\label{app:number}

\begin{figure}
    \centering
    \includegraphics[width=\linewidth]{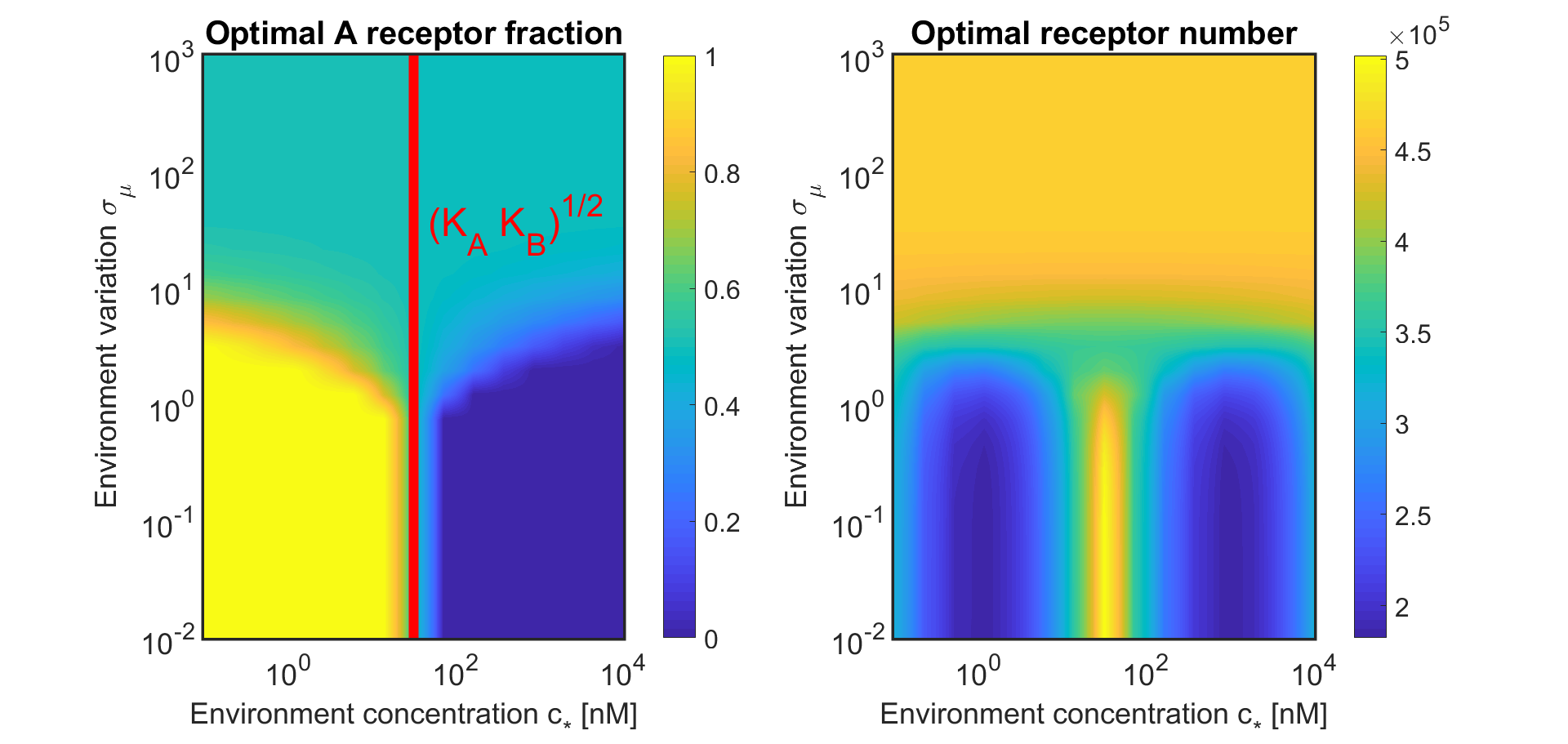}
    \caption{{ Transition between all-A and all-B is preserved in a model variant where the number of receptors is allowed to vary. Parameters are the same as Fig. 1 in the main paper, except for the penalty for increasing the number of receptors, which is $N^{penalty} = 50 N^{basal}$ (see text).}}
    \label{fig:float_nr}
\end{figure}

{Within the main text, we have followed earlier work in keeping the number of receptors on the cell fixed  \cite{hu2010physical,hu2011geometry,segota2013high,lakhani2017testing,fuller2010external,ueda2007stochastic,andrews2007information}. However, it is possible that when cells explore more complex environments, they should express different numbers of receptors depending on the typical concentration $c_*$ and the level of uncertainty $\sigma_\mu$. We address this possibility in \fig \ref{fig:float_nr}. }

Within the framework we have applied in this paper, accuracy always increases with increasing $N$ -- there are more measurements of the gradient, leading to increased accuracy (see \eq \ref{eq:variance_snapshot},\ref{eq:variance_naive},\ref{eq:variance_ERT} in the main text). If we allow the number of receptors to freely vary, and choose the number of receptors $N$ and receptor fractions $f$, we would find that the receptor number would increase without bound. This is obviously unphysical. Cells are under many restraints in controlling how many receptors they have, both in terms of the energetic cost of synthesizing them, and in the opportunity cost in taking up space on the cell surface.

In modeling cells with varying receptor number, we chose to find the receptor configuration that maximized $\cim \times e^{-N^{\textrm{added}}/N^{\textrm{penalty}}}$, where $N^{\textrm{added}}$ is the number of receptors expressed beyond the typical value $N^{\textrm{basal}} = 5 \times 10^4$, and $N^{\textrm{penalty}} = 50 N^{\textrm{basal}}$. This choice ensured that cells could easily express more than the basal level of receptors, but that expressing multiple orders of magnitude more receptors would be implausible -- consistent with the observed variation in receptor number on the membrane. We found that, though the optimal receptor numbers varied depending on the environment (\fig \ref{fig:float_nr}), the optimal receptor fractions closely agreed with those found assuming a constant number of receptors (\fig \ref{fig:hedge1}). 

Other choices for the penalty (e.g. optimizing $\cim + \alpha N^{\textrm{added}}$) gave different optimal receptor numbers but preserved the optimal receptor fractions and the transition between all-A, all-B, and the 50-50 mix. This suggests that the receptor fractions and the transition are highly robust to allowing the number of receptors to change. This may reflect that the optimal fractions are only very weakly dependent on the total number of receptors.

{Experimental measurements on Dictyostelium do see that receptors are internalized in response to saturating levels of chemoattractant; however, this happens on a long time scale ($\sim 5-10$ minutes) and results in a change of about 50\% of the receptors being internalized \cite{serge2011quantification,wang1988localization}. For Dictyostelium cells, which travel about a body length in a minute, we would expect that crawling cells would likely explore another concentration level $c_0$ before the receptor numbers adapt. Adaptation in eukaryotic chemotaxis is generally thought to occur on a post-receptor level \cite{tu2018adaptation,takeda2012incoherent}.}

\section{Extended data on hedging}
\label{app:extended}

\begin{figure*}[htb]
    \centering
    \includegraphics[width=\textwidth]{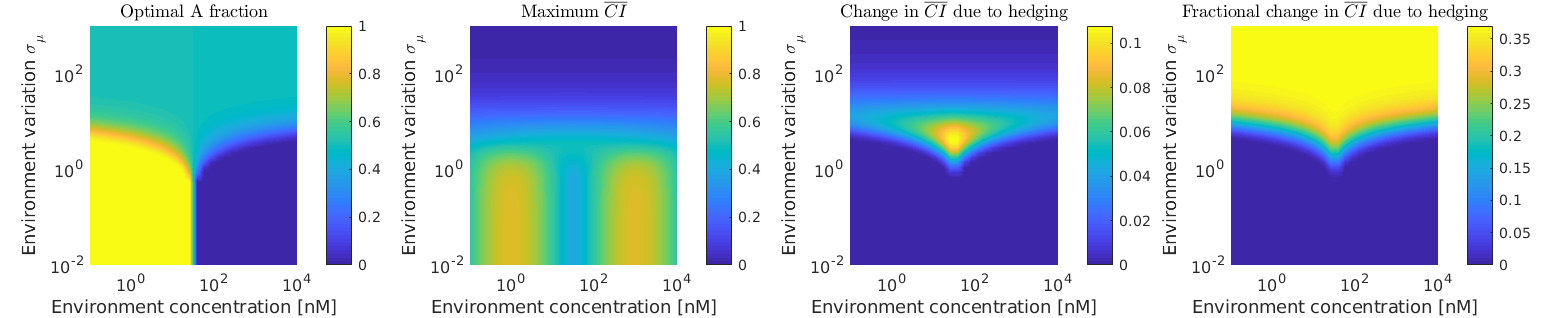}
    \caption{Tradeoffs in snapshot sensing. This figure complements Fig. 1 in the main text, showing how, with the same parameters, the maximum \cim depends on the uncertainty (2nd panel). The third and fourth panel show the increase in the mean CI due to hedging, i.e. the change vs all-$A$ or all-$B$ (whichever of these is better). The largest absolute improvements in \cim due to hedging are at intermediate uncertainties; in the limit of truly high uncertainties, no configuration creates a large \cim.}
    \label{fig:extra_hedging_snapshot}
\end{figure*}

\begin{figure*}[htb]
    \centering
    \includegraphics[width=\linewidth]{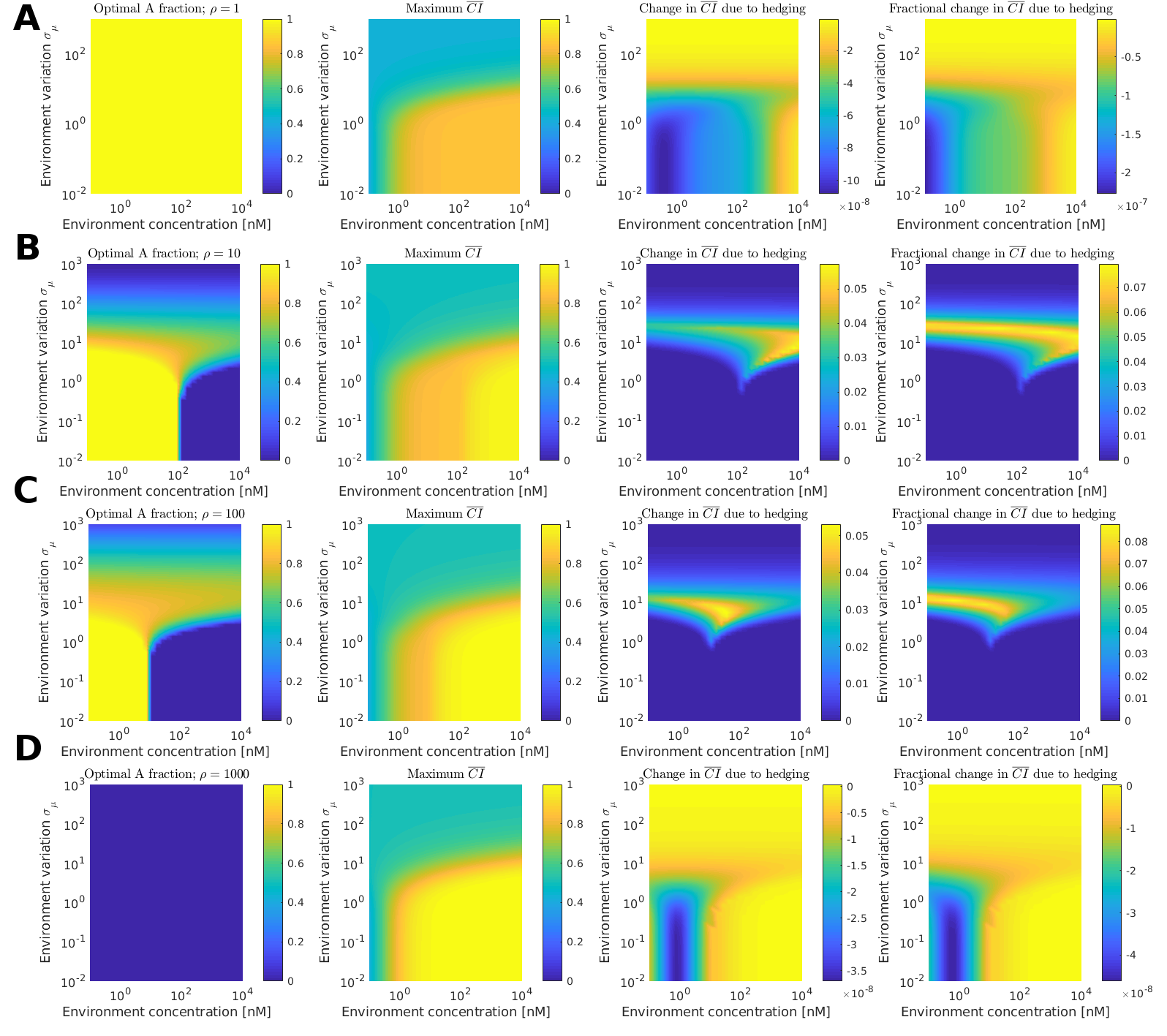}
    \caption{Tradeoffs in time-averaged sensing. This figure shows the maximum \cim and increase in \cim due to hedging for the time-average case. This corresponds to Fig. 4 in the main paper, with the left panels redrawing that data. We show these values for A) $\rho = 1$, B), $\rho = 10$, C) $\rho = 100$, D) $\rho = 1000$. Note that for A) and D), the change in mean CI due to hedging is slightly negative -- the optimal configuration is all-$A$ or all-$B$, but our optimization does not recover $f_A = 0,1$ with numerical precision.}
    \label{fig:extra_hedging_timeaverage}
\end{figure*}

Within the main paper, we have presented the optimal configuration of receptors as a function of the environment. However, at large environmental uncertainties, the benefit from hedging bets may not be as large. We show extended data corresponding to \fig 1 and \fig 4 in the main paper in \fig \ref{fig:extra_hedging_snapshot} and \fig \ref{fig:extra_hedging_timeaverage}.

\bibstyle{is-unsrt}
\bibliography{motility2}
\clearpage

\end{document}